\documentclass[conference]{IEEEtran}
\usepackage{etex}

\def\paperauthors{Michiel Van Beirendonck, Louis-Charles Trudeau, Pascal Giard, and Alexios Balatsoukas-Stimming}
\def\papertitle{A Lyra2 FPGA Core for Lyra2REv2-Based Cryptocurrencies}

\usepackage{ifpdf}
\ifpdf
\usepackage[pdfborder={0 0 0}]{hyperref}
\fi

\usepackage{cite}
\usepackage{graphicx}

\usepackage[cmex10]{amsmath}
\usepackage{amssymb}
\usepackage{nicefrac}
\usepackage{bm}

\usepackage{array}
\usepackage[caption=false,font=footnotesize]{subfig}

\usepackage{tabularx}
\usepackage{multirow}
\usepackage{multicol}
\usepackage{dcolumn}
\usepackage{booktabs}

\usepackage{fixltx2e}
\usepackage{stfloats}
\usepackage{algorithm}
\usepackage{algorithmicx}
\usepackage{algpseudocode}
\algrenewcommand\algorithmicindent{1.5em}%
\usepackage{float}

\usepackage{svg}

\usepackage{color}
\usepackage{tikz,pgf}
\usepackage{pgfplots}
\usetikzlibrary{shapes,positioning,arrows,decorations.markings,fit,calc,patterns,spy}

\usepackage{glossaries}
\usepackage{balance}

\usepackage[final]{changes}

\newcommand{\fixme}[2]{\ifx&#2&{\color{red}#1}\else{\color{red}FIXME\{}#1{\color{red}\}}\footnote{{\color{red}#2}}\PackageWarning{Fixme}{#1: #2}\fi}
\newcommand{\concat}{\:||\:}

\usepackage{xparse}
\makeatletter
\NewDocumentCommand{\LeftComment}{s m}{%
\Statex \IfBooleanF{#1}{\hspace*{\ALG@thistlm}}\(\triangleright\) #2}
\makeatother

\makeatletter
\NewDocumentCommand{\BlankComment}{s m}{%
\Statex \IfBooleanF{#1}{\hspace*{\ALG@thistlm}} #2}
\makeatother

\title{\papertitle}

\author{
	\IEEEauthorblockN{Michiel van Beirendonck\IEEEauthorrefmark{1}, Louis-Charles Trudeau\IEEEauthorrefmark{2}, Pascal Giard\IEEEauthorrefmark{2}, and Alexios Balatsoukas-Stimming\IEEEauthorrefmark{1}}
	\\
  \IEEEauthorblockA{
		\IEEEauthorrefmark{1}Telecommunications Circuits Laboratory, \'Ecole polytechnique f\'ed\'erale de Lausanne (EPFL), Lausanne, Switzerland\\
		\IEEEauthorrefmark{2}Electrical Engineering Department, \'Ecole de technologie sup\'erieure (\'ETS), Montr\'eal, Canada
  }
}
\begin{document}

\newacronym{kdf}{KDF}{key derivation function}
\newacronym{phs}{PHS}{password hashing scheme}
\newacronym{fsm}{FSM}{finite-state machine}
\newacronym[firstplural=D flip-flops (DFFs)]{dff}{DFF}{D flip-flop}
\newacronym{bram}{BRAM}{block RAM}
\newacronym{pow}{PoW}{proof-of-work}
\newacronym{cc}{CC}{clock cycle}
\newacronym{mux}{MUX}{multiplexer}

\maketitle

\begin{abstract}
  Lyra2REv2 is a hashing algorithm that consists of a chain of individual hashing algorithms and it is used as a \acrlong{pow} function in several cryptocurrencies that aim to be ASIC-resistant. The most crucial hashing algorithm in the Lyra2REv2 chain is a specific instance of the general Lyra2 algorithm. In this work we present the first FPGA implementation of the aforementioned instance of Lyra2 and we explain how several properties of the algorithm can be exploited in order to optimize the design.
\end{abstract}

\section{Introduction}
\label{sec:intro}
Recently, there has been a surge in the popularity of cryptocurrencies, which are digital currencies that enable transactions through a decentralized consensus mechanism. Most cryptocurrencies are based on a \emph{blockchain}, which is an ever-growing list of transactions that are grouped in blocks. Individual blocks in the chain are linked together using a cryptographic hash of the previous block, which ensures resistance against modifications, and every transaction is digitally signed. A blockchain needs to be protected from the double spending problem (i.e., an attacker spending the same digital money twice) and this is generally achieved by using a \glsfirst{pow} system. This system requires that new blocks provide proof that a certain amount of processing power went into constructing them before they get accepted in the chain. For cryptocurrencies, this is typically achieved by appending random numbers to a block until its cryptographic hash meets a certain condition. The chain with the most cumulative \gls{pow} is accepted as the correct one, so that an attacker must control more than half of the active processing power to perform a double-spend attack. Processing nodes that help to compute the hashes of new blocks (called \emph{miners}) are rewarded with a fraction of the cryptocurrency.

The first cryptocurrency, i.e., Bitcoin~\cite{bitcoin}, was initially mined using desktop CPUs. Then, GPUs were used to significantly increase the hashing speed. Eventually, GPU mining was outpaced by FPGA miners, which were in turn surpassed by ASIC miners. Nowadays, the majority of the computing power on the Bitcoin network is found in large ASIC farms, each operated by a single entity, which makes the decentralized nature of Bitcoin debatable. To solve this issue, new~\gls{pow} algorithms have been proposed that aim to be ASIC-resistant. ASIC resistance is achieved by using hashing algorithms that are highly serial, memory intensive, and parameterizable so that a manufactured ASIC can be easily made obsolete by simply changing some of the parameters, meaning that GPU mining is much more cost-effective. However, since GPUs are generally much less energy-efficient than ASICs, a massive adoption of ASIC-resistant cryptocurrencies would significantly increase the (already very high) energy consumption of cryptocurrency mining. FPGA-based miners, on the other hand, are flexible, energy efficient, and readily available to the general public at reasonable prices. Thus, they are an attractive platform for ASIC-resistant cryptocurrencies.

\begin{figure}[t]
  \centering
  \resizebox{0.9\columnwidth}{!}{
  \begin{tikzpicture}[node distance=0.4cm]
    \tikzstyle{hash} = [rectangle, minimum width=2.25cm, minimum height=0.6cm,text centered, draw=black]
    \tikzstyle{arrow} = [thick,->,>=stealth]

    \node (blake) [hash] {BLAKE-256};
    \node (keccak) [hash,right=of blake.east] {Keccak-256};
    \node (cube1) [hash,right=of keccak.east] {CubeHash-256};
    \node (lyra2) [hash,below=of cube1.south east] {\bf Lyra2};
    \node (skein) [hash,below=of lyra2.south west] {Skein-256};
    \node (cube2) [hash,left=of skein.west] {CubeHash-256};
    \node (bmw) [hash,left=of cube2.west] {BMW-256};

    \draw [arrow] ([xshift=-0.5cm]blake.west) -- (blake);
    \draw [arrow] (blake) -- (keccak);
    \draw [arrow] (keccak) -- (cube1);
    \draw [arrow] (cube1) -- (lyra2);
    \draw [arrow] (lyra2) -- (skein);
    \draw [arrow] (skein) -- (cube2);
    \draw [arrow] (cube2) -- (bmw);
    \draw [arrow] (bmw) -- ([xshift=-0.5cm]bmw.west);
  \end{tikzpicture}}\vspace*{-0.5em}%
  \caption{The Lyra2REv2 chained hashing algorithm.}
  \label{fig:Lyra2REv2}
\end{figure}
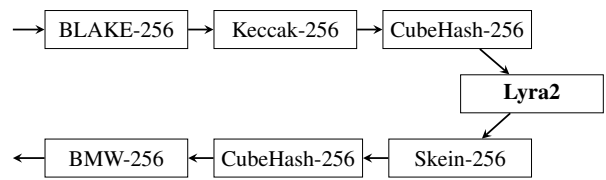

A prime example of an ASIC-resistant hashing algorithm is Lyra2REv2 (used by Vertcoin~\cite{vertcoin}, MonaCoin~\cite{monacoin}, and other cryptocurrencies), whose chained structure is shown in Fig.\,\ref{fig:Lyra2REv2}. The BLAKE, Keccak, Skein, BMW, and CubeHash hashing algorithms are well-known and have been studied heavily, both from a theoretical and from a hardware implementation perspective, as they were all candidates in the SHA-3 competition. On the other hand, to the best of our knowledge, no hardware implementation of the version of Lyra2~\cite{lyra2,lyra2TC} that is used in the Lyra2REv2 algorithm has been reported in the literature.

\subsubsection*{Contributions}
In this paper, we present the first hardware implementation of the version of Lyra2 used in Lyra2REv2 as a stepping stone towards the implementation of an energy-efficient FPGA miner for Lyra2REv2 cryptocurrencies. Post-layout results for two Xilinx FPGAs show that our proposed Lyra2 hardware architecture consumes very few FPGA resources to achieve a hashing throughput between 2.6~MHash/s and 3.7~MHash/s with an energy efficiency between \added{432}~nJ/Hash and \added{323}~nJ/Hash.

\section{Background}
\label{sec:bg}
In this section, we provide the necessary background on some components of the Keccak and BLAKE2 hashing algorithms, since they are also used in Lyra2.

\subsection{The Keccak Duplex}\label{sec:bg:keccak}
Keccak is a family of hashing algorithms based on a cryptographic sponge \cite{sponge,sha3}. A cryptographic sponge is a function that takes an arbitrary-length input to produce an arbitrary-length hashed output. Lyra2 uses a specific implementation of the sponge, called the \emph{duplex} construction, which has a state that is preserved across different inputs. The duplex construction with naming conventions as adopted in Lyra2 can be found in~\cite[Fig. 2]{lyra2refguide}. It consists of a permutation function $f$ that operates on a $w$-bit state vector, where $w=b+c$ and the parameters $b$ and $c$ are called the \emph{bitrate} and the \emph{capacity} of the sponge, respectively, as well as a padding rule \texttt{pad}. We note that the permutation $f$ is iterative and performs a pre-defined number of iterations, also called \emph{rounds}.

A call to the duplex construction proceeds as follows. An input string $M$ is first fed into the duplex. Then, it is padded to length $b$ and XOR'd into the lower $b$ bits of the state. The state is then fed through the permutation $f$. The output of $f$ is the new state of the duplex, while its lower $l$ bits are the output hash, where $l < b$. If we consider the duplex construction as an object $H$, then the aforementioned procedure is referred to as a method $H.\texttt{duplex}(M,l)$. The following two auxiliary methods are useful to simplify the notation: $H.\texttt{absorb}(M)$ updates the state using the input $M$ but discards the output (equivalent to $H.\texttt{duplex}(M,0)$), while $H.\texttt{squeeze}(l)$ reads $l$ output bits and then calls $H.\texttt{absorb}(\emptyset)$, where $\emptyset$ denotes an empty input string.

\subsection{The BLAKE2b Round Function}
BLAKE2 \cite{blake2} is a family of hash functions designed for fast software implementations. It is the successor of BLAKE as submitted to the SHA-3 competition \cite{blake}.
The Lyra2 algorithm heavily draws from the round function of BLAKE2b, the 64-bit variant of BLAKE2.
The round function consists of an arrangement of blocks that apply a so-called G-function to a 16-word state, where one G-function operates on 4 different state words.
For BLAKE2b a word has 64 bits meaning that 16 state words amount to 1024 bits. The total round transforms these 1024 bits using four G-blocks, rearranges the output, and then does a four G-block transformation again. Algorithm\,\ref{algo:blake2b:g} describes the modified BLAKE2b G-function as used in Lyra2.

\begin{algorithm}[t]
  \caption{The G-function of BLAKE2b as used in Lyra2}
  \label{algo:blake2b:g}
  \begin{algorithmic}[1]\small
    \State\texttt{INPUTS:} $a,b,c,d$
    \State \texttt{OUTPUTS:} $a',b',c',d'$
    \State $a'\gets a + b$
    \State $d'\gets (d \oplus a') \ggg 32$
    \State $c'\gets c + d'$
    \State $b'\gets (b \oplus c') \ggg 24$
    \State $a'\gets a' + b'$
    \State $d'\gets (d' \oplus a') \ggg 16$
    \State $c'\gets c' + d'$
    \State $b'\gets (b' \oplus c') \ggg 63$
  \end{algorithmic}
\end{algorithm}

\section{The Simplified Lyra2 Algorithm of Lyra2REv2}

\begin{algorithm}
  \caption{The Lyra2 algorithm as specified in Lyra2REv2.}
  \label{algo:lyra2}
  \begin{algorithmic}[1]\fontsize{8.75pt}{8.75pt}\selectfont
   	\State \texttt{PARAMS:} $H, \rho, \omega, T, R, C, k, b\text{ as }H.b$
    \State\texttt{INPUT:} $pwd$
    \State \texttt{OUTPUT:} $K$

    \vspace*{0.3em}
    \LeftComment{\textbf{Bootstrapping Phase}}

    \State $params \gets \texttt{len}(K) \concat \texttt{len}(pwd) \concat \texttt{len}(pwd) \concat T \concat R \concat C$
    \State $H.\texttt{absorb}(\texttt{pad}(pwd \concat pwd \concat params))$  \label{lst:line:boostrap}

    \vspace*{0.3em}
    \LeftComment{\textbf{Setup Phase}}
    \For{$col \gets 0$ to $C - 1$} \label{lst:line:setup0}
    \State $M[0][C-1-col] \gets H_{\rho}\texttt{.squeeze}(b)$
    \EndFor \label{lst:line:setup0end}

    \For{$col \gets 0$ to $C - 1$}  \label{lst:line:setup1}
    \State $M[1][C-1-col] \gets M[0][col] \oplus H_\rho.\texttt{duplex}(M[0][col],b)$
    \EndFor \label{lst:line:setup1end}

    \For{$row^0 \gets 2$ to $R-1$} \label{lst:line:setup2}
    \State $prev^0 \gets row^0 - 1$
    \State $row^1 \gets row^0 - 2$
    \For{$col \gets 0$ to $C - 1$}
    \State $rand \gets H_\rho.\texttt{duplex}(M[row^1][col] \boxplus M[prev^0][col],b)$
    \State $M[row^0][C-1-col] \gets M[prev^0][col] \oplus rand$
    \State $M[row^1][col] \gets M[row^1][col] \oplus (rand \lll \omega)$
    \EndFor
    \EndFor \label{lst:line:setup2end}

    \vspace*{0.3em}
    \LeftComment{\textbf{Wandering Phase}}
    \For{$row^0 \gets 0$ to $R \cdot T-1$} \label{lst:line:wander}
    \State $prev^0 \gets row^0 - 1$

    \State $row^1 \gets \texttt{lsw}(rand) \texttt{ mod } R$
    \For{$col \gets 0$ to $C - 1$}
    \fontsize{8.75pt}{3.0pt}\selectfont\State $rand \gets H_\rho.\texttt{duplex}(M[row^1][col] \boxplus M[prev^0][col],b)$
    \small
    \State $M[row^0][col] \gets M[row^0][col] \oplus rand$ \label{lst:line:collide1}
    \State $M[row^1][col] \gets M[row^1][col] \oplus (rand \lll \omega)$ \label{lst:line:collide2}
    \EndFor
    \EndFor

    \vspace*{0.3em}
    \LeftComment{\textbf{Wrap-up Phase}} \label{lst:line:wrap}
    \State $H.\texttt{absorb}(M[row^1][0])$
    \State $K \gets H.\texttt{squeeze}(k)$

  \end{algorithmic}
\end{algorithm}

Lyra2 was initially created as a \gls{phs} for secure storage \cite{lyra2,lyra2TC}.

Lyra2 uses the duplex construction from Keccak, where the permutation function $f$ is the round function from BLAKE2b.
In the remainder of the text, calls to a full-round (i.e., 12 iterations) duplex will be denoted as calls to $H$, while reduced-round duplexing as calls to $H_\rho$, where $\rho$ denotes the reduced number of rounds. Because the G-functions are specified to operate on an array of 16 64-bit words, Lyra2 uses a duplex with a width of $w = 16 \cdot 64 = 1024$\,bits. Pseudocode for the simplified version of Lyra2 that is used specifically in Lyra2REv2 is given in Algorithm\,\ref{algo:lyra2} and can be compared to the original Lyra2 pseudocode available in~\cite{lyra2}. In the following sections, we explain each phase of the simplified Lyra2 algorithm in more detail.


\subsection{Bootstrapping Phase}
In the bootstrapping phase, the duplex is initialized with a state that depends on the password input $pwd$, a salt (which in Lyra2REv2 is set to be equal to $pwd$), and the parameters $T$, $R$, and $C$ by using a full-round absorb. The duplex $H$ in Algorithm \,\ref{algo:lyra2} internally uses a bitrate $b = 768$ bits and a capacity $c = 256$ bits. The $H.\texttt{absorb}(\cdot)$ call on line \,\ref{lst:line:boostrap}, however, considers only inputs of 512 bits instead of $b$ bits, so as to not overwrite the upper part of the initialization state, i.e, the 512-bit initialization value \textit{IV} specified by BLAKE2b. This results in two full-round absorbs, where the first and second absorbs process $(pwd \concat pwd)$ and \texttt{pad}$(params)$, respectively.

\subsection{Setup Phase}
During the setup phase, an $R \times C \times b$ memory matrix $M$ is initialized using the single-round duplex $H_{1}$. During setup, rows are initialized from first to last, while columns within a row are initialized from last to first. From the second row onward a previous row is re-read, making it impractical to only store parts of the memory matrix. Also, from the third row onward, in addition to the previous row, i.e., $prev^0$, a specific pre-initialized row, i.e., $row^1$, is revisited (i.e., read and updated) in a deterministic manner. Rows are re-read or revisited from the first to the last column. Revisited rows use a rotated version of the duplex output, where the rotation number is chosen as $\omega = 64$ in Lyra2REv2.

\subsection{Wandering Phase}
The wandering phase is generally the most time-consuming phase and it proceeds similarly to the setup phase. Specifically, it revisits two rows $row^0$ and $row^1$, where $row^0$ is chosen deterministically but $row^1$ is chosen in a pseudorandom fashion by using the least significant part of the duplex output. We note that the pseudorandom and deterministic row can collide, resulting in the operations on line \,\ref{lst:line:collide1} and \,\ref{lst:line:collide2} to sequentially read from and then write to the same matrix cell.

\subsection{Wrap-up Phase}
The wrap-up phase consists of a full-round absorb of a specific cell of $M$ followed by a squeeze of the hashed output $K$. This specific cell is likewise pseudorandom, as it is selected as the first cell of the lastly revisited pseudorandom row. The requested squeeze length $k = 256$ is lower than the bitrate $b=768$, which means that the final output is provided directly from the duplex state without a permutation $f$.

\section{FPGA Implementation of Simplified Lyra2}\label{sec:impl}

In the current instance of Lyra2 as used in Lyra2REv2, the timecost parameter is $T=1$, the number of rows in the memory matrix is $R=4$, the number of columns in the memory matrix is $C=4$, and the desired hashing output length is $k=256$. We note that our architecture is optimized for these parameter values, but it can be modified relatively easily to accommodate potential parameter changes if a hard fork is decided. Moreover, for the aforementioned parameters, the memory matrix $M$ is 1.5~kB in size, which is clearly not prohibitively large to be implemented either on an FPGA or on an ASIC. 
\added{The claimed ASIC-resistance of the Lyra2REv2 algorithm comes from the fact that $T$, $C$, and $R$ can be increased easily if necessary.}

\begin{figure}[t]
  \centering
  \includegraphics[width=0.8\columnwidth]{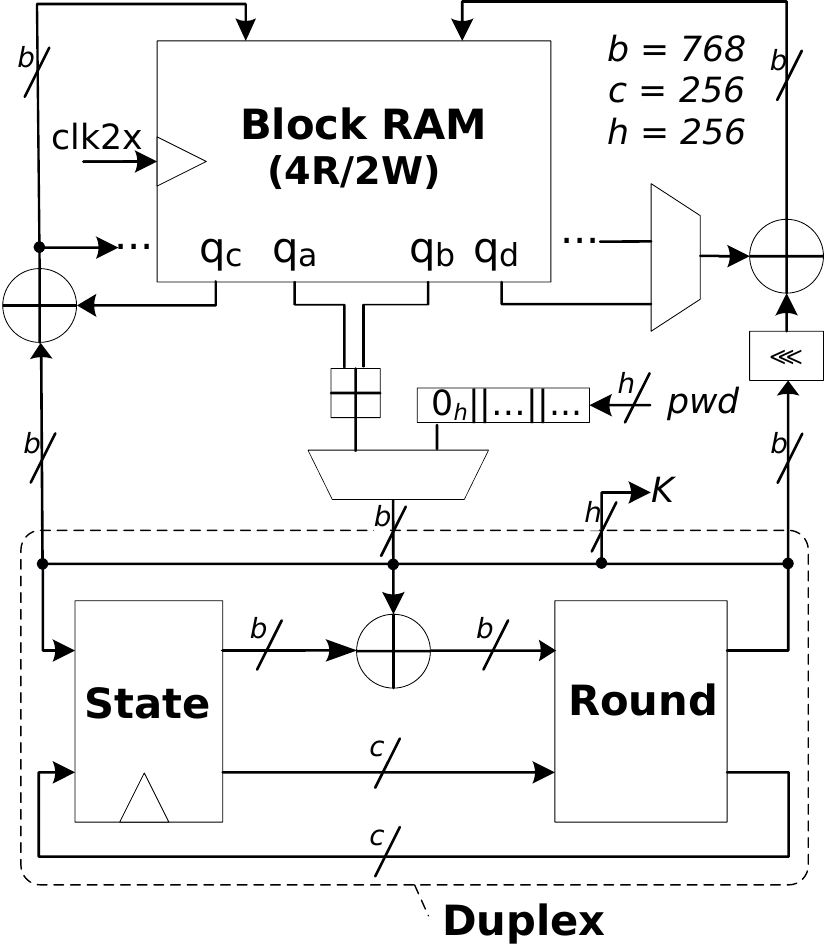}%
  \vspace{-0.75em}%
  \caption{Datapath of our proposed Lyra2 FPGA architecture.}
  \label{fig:lyra2-data}%
  \vspace{-0.75em}
\end{figure}

The datapath of our proposed FPGA implementation of the simplified Lyra2 algorithm used in Lyra2REv2 is shown in Fig.\,\ref{fig:lyra2-data},
where the duplex construction with its state, round, and XOR input block can be clearly distinguished. The memory matrix $M$ is mapped to a \gls{bram}.
\added{To reduce the complexity of the \gls{mux} at the input of the duplex, the \gls{bram} also contains constant vectors of $b$ bits used during the bootstrapping and setup phases: an all-zero vector and the $\texttt{pad}(params)$ vector.}
We first describe a version of the hardware architecture where each round of the $f$ function is executed in a single \gls{cc}. We then describe how this basic architecture can be improved through pipelining.

\subsection{Basic Iterative Architecture}

Our basic iterative Lyra2 architecture requires 68 \glspl{cc} per hash: 24 for the bootstrapping phase, 16 for the setup and wandering phases, and 12 for the wrap-up phase.

\subsubsection{Bootstrapping Phase} During the bootstrapping phase, the duplex processes two 512-bit input blocks from $\texttt{pad}(pwd \concat pwd \concat params)$ using a full-round absorb. In Lyra2REv2, $pwd = cube_{\text{out}}$, with $cube_{\text{out}}$ the output from CubeHash, the previous algorithm in the chain. 
\added{Thus, as shown in Fig.\,\ref{fig:lyra2-data}, the $(pwd \concat pwd)$ vector is one of the inputs to the duplex's \gls{mux}. On the other hand, the $\texttt{pad}(params)$ vector is fed into the sponge by loading it on $q_a$ while simultaneously loading the all-zero vector on $q_b$. Both constants are stored at known addresses in the \gls{bram}, and are absorbed in a separate 12-round \texttt{Bootstrap} state. During bootstrapping, the duplex only receives an input vector in the first round. Hence, for subsequent rounds, $q_a$ and $q_b$ output the all-zero vector, and their sum is passed to the duplex via its input \gls{mux}.}

\subsubsection{Setup Phase} We split the setup phase into three distinct phases for convenience, namely \texttt{Setup0}, \texttt{Setup1}, and \texttt{Setup2}, which correspond to Lines~\ref{lst:line:setup0}--\ref{lst:line:setup0end}, Lines~\ref{lst:line:setup1}--\ref{lst:line:setup1end}, and Lines~\ref{lst:line:setup2}--\ref{lst:line:setup2end} of Algorithm~\ref{algo:lyra2}, respectively. \added{Similarly to the bootstrapping phase, the setup phase uses the all-zero vector stored in the \gls{bram}. In the \texttt{Setup0} state, the squeezes input an empty message into the duplex and directly write the duplex output to the \gls{bram}. To achieve that, the all-zero vector is output on $q_a$, $q_b$, and $q_c$. \texttt{Setup1} reads the all-zero vector on $q_b$, but a specific vector from the \gls{bram} on $q_a$. \texttt{Setup2} reads two vectors from $q_a$ and $q_b$. Both the duplex output and the rotated duplex output are XORed with two other vectors from the \gls{bram}, requiring the two XOR blocks in parallel illustrated in Fig.\,\ref{fig:lyra2-data}.} On the control path, counters keep track of the various rows ($row^0, row^1, prev^1$) and their corresponding columns to generate read and write addresses for the RAM.

\subsubsection{Wandering Phase} The input to the duplex in the wandering phase is always the word-wise addition of two RAM cells. Both XOR blocks connected to the duplex output are used.
As mentioned before, the pseudorandom and deterministic rows used in the wandering phase can collide. In hardware, this special case requires the output of one XOR block to input to the other, while the write port of the first XOR block needs to be disabled to prevent write collisions on the RAM.

\subsubsection{Wrap-Up Phase} Wrap-up inputs one RAM cell into the sponge and then processes it using a full-round absorb. For the following squeeze, the requested hashed-output length $k$ is lower than the bitrate $b$, meaning that the duplex state at that point directly provides the output hash. 

\subsection{Memory Matrix}
In the wandering phase, up to two RAM cells need to be written and three RAM cells need to be read per \gls{cc}. These operations cannot be spread over multiple \glspl{cc} without affecting the overall throughput of the design. Therefore, we use standard two-port \glspl{bram} along with multipumping and replication techniques~\cite{Laforest2010} in order implement the required functionality.
Replication provides extra read ports by physically replicating the \gls{bram} while connecting the write ports to keep the two copies coherent. Multipumping operates the \gls{bram} at double the clock frequency of the surrounding logic, which, together with replication, effectively provides four read ports and two write ports.

\subsection{Pipelined Architecture}

Pipelining the BLAKE2b round function can greatly reduce the delay of the critical path, which extends from the RAM read ports to the RAM write ports in the basic iterative version described above. \added{Eight} pipeline stages in the round were found to optimally increase throughput/area.
Each hash that is concurrently being processed by the core needs its own memory.
However, extra RAM-based memory is readily available since the current Lyra2REv2 parameters result in a RAM depth much shallower than that of the FPGA \glspl{bram}.
With adequate scheduling, concurrent hashes write to the same \glspl{bram} in distinct \glspl{cc}.
While read ports $q_a$ and $q_b$ feed the duplex, $q_c$ and $q_d$ feed the XORs with duplex outputs. When pipelining the round function, $q_c$ and $q_d$ therefore need to be delayed by as many \glspl{cc} as there are pipeline stages. The extra read port that is unused in the basic architecture allows delaying the control path for $q_d$ rather than using a delayed version of $q_b$, avoiding a long chain of 768-bit registers.
\added{Eight} pipeline stages in the round increase the latency to \added{544} \glspl{cc} per hash. On the other hand, \added{eight} hashes are processed concurrently and the achievable clock frequency more than doubles, so the overall hashing throughput is improved significantly.

\section{Hardware Implementation Results}
\added{To the best of our knowledge, there is no FPGA implementation of Lyra2 in the literature. For this reason, we can unfortunately not provide comparative FPGA implementation results. Moreover, since in this work we only present a core for Lyra2 and not a full miner architecture, we can also not compare our implementation to existing FPGA miners for cryprocurrencies based on other \gls{pow} algorithms.}

Table\,\ref{tab:impl:results} presents post-fitting results for the Xilinx Virtex 7 485T FPGA featured on the Xilinx VC707 Evaluation Kit as well as for the Xilinx Zynq Ultrascale+ 7EV FPGA from the ZCU104 Evaluation Kit.
\added{The power-consumption estimation was obtained using Xilinx's Vivado Power Estimator tool, where the timing constraints are those required for the operating frequencies of Table\,\ref{tab:impl:results}, the switching activity is from the simulation of the Lyra2 core processing random input vectors, and the post-fitted design provided to the tool meets all timing constraints. Table\,\ref{tab:impl:results} reports the estimated dynamic power for the Lyra2 core.}
\added{The functionality of the Lyra2 core was verified against test vectors that were generated using CPUminer~\cite{cpuminer}.}

From Table\,\ref{tab:impl:results}, looking at the number of slices or CLBs required for the Virtex and Zynq FPGAs, respectively, it can be seen that the proposed Lyra2 core amounts to less than \added{4\%} of the resources available. The amount of RAM occupied is the same for both FPGAs, however the usage share is greater for the Zynq as it has less RAM blocks than FPGAs from the Virtex series. Also, from Table\,\ref{tab:impl:results}, it can be seen that the throughout is 2.58\,MHash/s and 3.69\,MHash/s for the Virtex and Zynq FPGAs, respectively. The estimated dynamic power consumption of the Lyra2 core is under 1.2\,W for both FPGAs. As a result, the energy efficiency is estimated to be in the vicinity of \added{325} to \added{435}\,nJ/Hash.

\begin{table}[t]
  \centering
  \caption{Post-fitting results for Xilinx FPGAs.}
  \begin{tabular}{l | c | c}
    \toprule\hline
    \multirow{2}{*}{\bf FPGA} & Virtex 7 & Zynq\\
    & 485T & Ultrascale+ 7EV \\
    \hline\hline
    Area (slices or CLBs)& 2\,163 (2.85\%) & 1\,153 (4.00\%) \\
    ~~~LUTs              & 6\,047 (1.99\%) & 6\,010 (2.61\%) \\
    ~~~Registers         & 8\,296 (1.37\%) & 8\,296 (1.80\%) \\
    ~~~RAM (kbits)       & 1\,548 (4.27\%) & 1\,548 (5.39\%)\\\hline
    Frequency (MHz)      & 175 & 250 \\
    T/P (MHash/s)        & 2.58 & 3.69 \\
    Dyn. Power (W)       & 1.12 & 1.19 \\
    Energy Eff. (nJ/Hash)& 432 & 323 \\
    \hline
    \bottomrule
  \end{tabular}
  \label{tab:impl:results}
\end{table}

\section{Conclusion}\label{sec:conclusion}
In this paper, we presented the first hardware implementation of the Lyra2 hashing algorithm, tailored to Lyra2REv2, an ASIC-resistant chained hashing algorithm employed by a few cryptocurrencies. The key to achieve good throughput and energy efficiency is to efficiently map the memory matrix to FPGA RAM blocks and to pipeline the BLAKE2b round function. Based on post-fitting results for two Xilinx FPGAs, we believe that the proposed Lyra2 implementation is a promising core for the purpose of FPGA-based Lyra2REv2 mining.\footnote{Our VHDL code and relevant scripts are publicly available at\\\url{https://github.com/Michielvb/lyra2-hw}.} For example, we showed that, for a Zynq Ultrascale+ FPGA featured on an affordable evaluation kit, the achievable throughput is of 3.7\,MHash/s and the energy efficiency of \added{323}\,nJ/Hash, for a resource usage of \added{4}\%.

\section*{Acknowledgment}
The authors thank Jean-Franc\c{}ois T{\^e}tu for useful feedback. The authors also gratefully acknowledge the support of NVIDIA Corporation with the donation of a Titan Xp GPU. 
\balance
\bibliographystyle{IEEEtran}
\bibliography{IEEEabrv,ConfAbrv,refs}

\begin{thebibliography}{10}
\providecommand{\url}[1]{#1}
\csname url@samestyle\endcsname
\providecommand{\newblock}{\relax}
\providecommand{\bibinfo}[2]{#2}
\providecommand{\BIBentrySTDinterwordspacing}{\spaceskip=0pt\relax}
\providecommand{\BIBentryALTinterwordstretchfactor}{4}
\providecommand{\BIBentryALTinterwordspacing}{\spaceskip=\fontdimen2\font plus
\BIBentryALTinterwordstretchfactor\fontdimen3\font minus
  \fontdimen4\font\relax}
\providecommand{\BIBforeignlanguage}[2]{{%
\expandafter\ifx\csname l@#1\endcsname\relax
\typeout{** WARNING: IEEEtran.bst: No hyphenation pattern has been}%
\typeout{** loaded for the language `#1'. Using the pattern for}%
\typeout{** the default language instead.}%
\else
\language=\csname l@#1\endcsname
\fi
#2}}
\providecommand{\BIBdecl}{\relax}
\BIBdecl

\bibitem{bitcoin}
S.~Nakamoto, ``Bitcoin: A peer-to-peer electronic cash system,'' 2008.

\bibitem{vertcoin}
\BIBentryALTinterwordspacing
``Vertcoin.'' [Online]. Available: \url{http://vertcoin.org}
\BIBentrySTDinterwordspacing

\bibitem{monacoin}
\BIBentryALTinterwordspacing
``Mona{C}oin.'' [Online]. Available: \url{https://monacoin.org}
\BIBentrySTDinterwordspacing

\bibitem{lyra2}
\BIBentryALTinterwordspacing
M.~A. Simpl{\'\i}cio~Jr, L.~C. Almeida, E.~R. Andrade, P.~C. dos Santos, and
  P.~S. Barreto, ``Lyra2: Password hashing scheme with improved security
  against time-memory trade-offs,'' Cryptology ePrint Archive, Report 2015/136,
  2015. [Online]. Available: \url{https://eprint.iacr.org/2015/136}
\BIBentrySTDinterwordspacing

\bibitem{lyra2TC}
E.~R. Andrade, M.~A. Simplicio, P.~S. L.~M. Barreto, and P.~C.~F. d.~Santos,
  ``Lyra2: Efficient password hashing with high security against time-memory
  trade-offs,'' \emph{{IEEE} Trans. Comput.}, vol.~65, no.~10, pp. 3096--3108,
  Oct 2016.

\bibitem{sponge}
G.~Bertoni, J.~Daemen, M.~Peters, and G.~V. Assche, ``Cryptographic sponge
  functions,'' Tech. Report v0.1, Jan. 2011.

\bibitem{sha3}
{NIST}, ``{SHA-3} standard: Permutation-based hash and extendable output
  functions,'' {FIPS} Publication 202, Aug. 2015.

\bibitem{lyra2refguide}
M.~A. Simplicio~Jr, L.~C. Almeida, E.~R. Andrade, P.~C. dos Santos, and P.~S.
  Barreto, ``The {Lyra2} reference guide,'' Tech. Report v2.3.2, 2014.

\bibitem{blake2}
J.-P. Aumasson, S.~Neves, Z.~Wilcox-O'Hearn, and C.~Winnerlein, ``{BLAKE2}:
  simpler, smaller, fast as {MD5},'' in \emph{Int. Conf. on Applied Crypto. and
  Netw. Security (ACNS)}.\hskip 1em plus 0.5em minus 0.4em\relax Springer,
  2013, pp. 119--135.

\bibitem{blake}
J.-P. Aumasson, L.~Henzen, W.~Meier, and R.~C.-W. Phan, ``{SHA-3} proposal
  {BLAKE},'' Tech. Report v1.3, Dec. 2010.

\bibitem{Laforest2010}
C.~E. LaForest and J.~G. Steffan, ``Efficient multi-ported memories for
  {FPGAs},'' in \emph{Ann. ACM/SIGDA Int. Symp. on FPGAs}, 2010, pp. 41--50.

\bibitem{cpuminer}
\BIBentryALTinterwordspacing
T.~Pruvot, ``{CPUMiner-Multi},'' GitHub repository, 2017. [Online]. Available:
  \url{https://github.com/tpruvot/cpuminer-multi}
\BIBentrySTDinterwordspacing

\end{thebibliography}

\end{document}